\documentclass[twocolumn]{aastex631}

\newcommand{\kms}{\mathrm{km\,s}^{-1}}

\usepackage{amsmath}	
\usepackage{amssymb}	

\begin{document}

\title{A Massive Yellow Supergiant in the Far Outer Disk of M31: Evidence for In Situ Massive Star Formation Beyond the Optical Radius}

\author[0009-0007-5623-2475]{Pinjian Chen}
\affiliation{National Astronomical Observatories, Chinese Academy of Sciences, Beijing 100101, P.\,R.\,China}
\affiliation{School of Astronomy and Space Science, University of the Chinese Academy of Sciences, Beijing, 100049, P.\,R.\,China}

\author[0000-0003-2472-4903]{Bingqiu Chen}
\affiliation{South-Western Institute for Astronomy Research, Yunnan University, Kunming, Yunnan 650091, P.\,R.\,China}

\author[0000-0003-2471-2363]{Haibo Yuan}
\affiliation{School of Physics and Astronomy, Beijing Normal University, Beijing 100875, P.\,R.\,China}
\affiliation{Institute for Frontiers in Astronomy and Astrophysics, Beijing Normal University, Beijing 102206, P.\,R.\,China}

\author[0000-0003-1286-2743]{Xuan Fang}
\affiliation{National Astronomical Observatories, Chinese Academy of Sciences, Beijing 100101, P.\,R.\,China}
\affiliation{School of Astronomy and Space Science, University of the Chinese Academy of Sciences, Beijing, 100049, P.\,R.\,China}
\affiliation{Xinjiang Astronomical Observatory, Chinese Academy of Sciences, 150 Science 1-Street, Urumqi, Xinjiang, 830011, P.\,R.\,China}

\author[0000-0001-7084-0484]{Xiaodian Chen}
\affiliation{National Astronomical Observatories, Chinese Academy of Sciences, Beijing 100101, P.\,R.\,China}
\affiliation{School of Astronomy and Space Science, University of the Chinese Academy of Sciences, Beijing, 100049, P.\,R.\,China}
\affiliation{Institute for Frontiers in Astronomy and Astrophysics, Beijing Normal University, Beijing 102206, P.\,R.\,China}

\author[0000-0002-9390-9672]{Chao-Wei Tsai}
\affiliation{National Astronomical Observatories, Chinese Academy of Sciences, Beijing 100101, P.\,R.\,China}
\affiliation{School of Astronomy and Space Science, University of the Chinese Academy of Sciences, Beijing, 100049, P.\,R.\,China}
\affiliation{Institute for Frontiers in Astronomy and Astrophysics, Beijing Normal University, Beijing 102206, P.\,R.\,China}

\author[0000-0003-0985-6166]{Kai Zhang}
\affiliation{National Astronomical Observatories, Chinese Academy of Sciences, Beijing 100101, P.\,R.\,China}
\affiliation{School of Astronomy and Space Science, University of the Chinese Academy of Sciences, Beijing, 100049, P.\,R.\,China}

\author[0000-0003-1295-2909]{Xiaowei Liu}
\affiliation{South-Western Institute for Astronomy Research, Yunnan University, Kunming, Yunnan 650091, P.\,R.\,China}

\correspondingauthor{Bingqiu Chen}
\email{bchen@ynu.edu.cn}



\begin{abstract}
While massive stars are known to shape galactic ecosystems, their formation has long been assumed to require the high-density environments of inner galactic disks. This paradigm is challenged by mounting evidence of young massive stars in extended galaxy outskirts, yet direct confirmation of in situ massive star formation in such extreme low-density environments remains scarce. Here, we present the discovery of LAMOST J0048+4154, a massive yellow supergiant situated at a deprojected galactocentric distance of $\sim$34\,kpc in M31, making it the most distant massive star confirmed in this galaxy. Through spectroscopic and photometric analyses, we classify J0048+4154 as an F5--F8I supergiant with an effective temperature of $6357^{+121}_{-118}$\,K and a luminosity of $\log L/L_{\odot} = 5.00^{+0.06}_{-0.06}$, corresponding to an $\sim$18\,$M_{\odot}$ progenitor and an age of $\sim$10\,Myr. FAST H\,\textsc{i} observations reveal close spatial and kinematic alignment between the star and a faint H\,\textsc{i} external arm, suggesting in situ formation in a region of low gas density. The presence of other UV-bright, early-type stars in the vicinity further supports low-level recent star formation in M31's very outer disk. These findings challenge the prevailing assumption that massive star formation is confined to inner disks or classical star-forming regions and underscore the need to re-examine the role of spiral galaxy outskirts in fueling and sustaining star formation. J0048+4154 thereby expands our understanding of the extent of M31's young stellar component and exemplifies how outer disks may harbor conditions conducive to forming massive stars, despite low-density environments.
\end{abstract}



\section{Introduction} \label{sec:intro}

The outer regions of spiral galaxies offer valuable insight into how disk dynamics, gas accretion, and star formation interact in low-density environments. Although galaxy outskirts typically exhibit low gas surface densities and metallicities, young stellar populations have been identified in the extreme outskirts of the Milky Way and other spiral galaxies \citep{Lopez2018, Jang2020, Koda2022, Zhang2025}. In the Andromeda Galaxy (M31), the nearest massive spiral to the Milky Way, observations beyond $\sim$20\,kpc show a significant decline in star formation efficiency, with older stellar populations dominating these extended regions, as revealed by optical and near-infrared surveys \citep{Ferguson2002, Gilbert2012}. Despite this, recent H\,\textsc{i} observations have uncovered filamentary gas structures in the outskirts of M31, indicating possible ongoing gas accretion or remnants of tidal interactions \citep{Braun2009, Chemin2009, Kerp2016}. Theoretical studies suggest that such structures may locally compress gas and trigger star formation, even in environments otherwise deemed hostile to stellar birth \citep{Bournaud2007, Elmegreen2010}. Yet, direct evidence of massive star formation in these outermost regions remains limited.

\begin{figure*}
\centering 
\includegraphics[width=2.1\columnwidth]{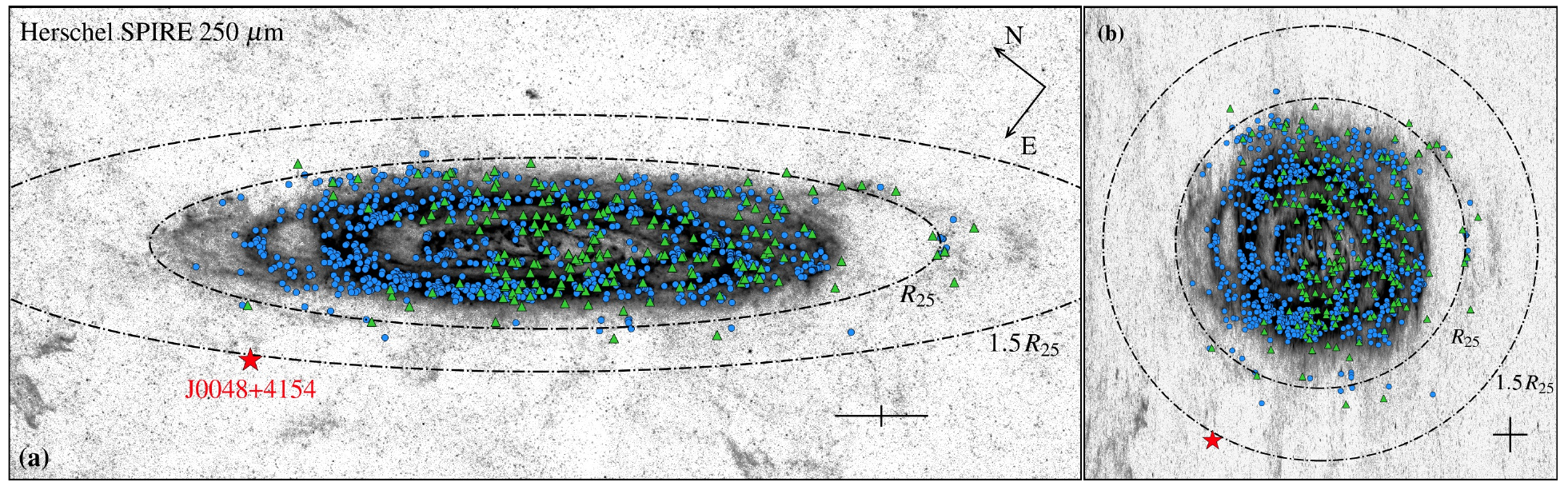}
\caption{
Location of J0048+4154 in M31. 
(a) \textit{Herschel} SPIRE 250\,$\mu$m image displayed on a logarithmic scale. Green triangles mark supergiant candidates identified from LAMOST, selected based on the sample of \citet{Chen2025} with $P_{\mathrm{M31}}>0.8$ and $P_{\mathrm{MW}}<0.1$, where $P_{\mathrm{M31}}$ and $P_{\mathrm{MW}}$ refer to the probability that the star belongs to M31 and the Milky Way, respectively. Blue circles indicate spectroscopically confirmed massive stars compiled from the literature \citep{Maravelias2022}. The dot-dashed ellipses correspond to galactocentric radii of $R_{\rm GC} = R_{25}$ and $1.5R_{25}$. The black cross denotes a physical scale of 5\,kpc in both the $x$- and $y$-directions. 
(b) Same image as (a), but with the $y$-axis deprojected assuming a fixed inclination angle of 77.5\arcdeg. 
\label{fig: H250}}
\end{figure*}

In this letter, we report the discovery of a massive yellow supergiant (YSG), LAMOST\,J004838.60+415420.7 (hereafter J0048+4154), with an estimated mass of 18\,M$_\odot$, located at an unexpected galactocentric distance of $\sim$34\,kpc (Fig.~\ref{fig: H250}). This finding challenges conventional notions of star formation thresholds in low-density environments and offers new insight into the mechanisms that may enable massive star formation far from galactic centers. The origin of  the progenitor cloud of this star raises important questions: whether it formed in situ from accreted gas, emerged during a past merger-driven starburst, or migrated dynamically from the inner disk. Research on M31's outer disk has largely focused on its stellar halo substructures \citep{Ibata2014} and diffuse H\,\textsc{i} kinematics \citep{Chemin2009}, leaving the star formation processes in these regions poorly understood. By combining stellar population data with insights into the interstellar medium (ISM) and chemical enrichment, our study highlights the need to revisit models of gas cycling and feedback in galaxy outskirts. This case also provides a valuable benchmark for cosmological simulations, which often struggle to reproduce extended star-forming disks without invoking high gas accretion rates \citep[e.g.,][]{Tissera2016}.

Throughout this paper, we adopt a distance to M31 of $785 \pm 25$\,kpc \citep{McConnachie2005}, corresponding to a linear scale of approximately 13.7\,kpc\,deg$^{-1}$. The optical radius is taken as $R_{25} = 95.3\arcmin$ \citep{deVaucouleurs1991}. We assume the disk center of M31 at coordinates (00$^{\rm h}$42$^{\rm m}$44.33$^{\rm s}$, +41$^\circ$16\arcmin07.5\arcsec), with an inclination of 77.5\arcdeg\ \citep{Simien1978} and a position angle of 37.7\arcdeg\ \citep{Haud1981}.

\begin{table*}
\caption{Summary of Spectroscopic Observations\label{tab: observations}}
\setlength{\tabcolsep}{13.pt}
\begin{center}
\begin{tabular}{cccccccc}
\hline
\hline
Date  & Telescope/ & Wavelength   & FWHM  & Exposure & Seeing & S/N \\
 (UT)  & Instrument & Range (\AA) & (\AA) &  (s) & (\arcsec) & \\

\hline
2011-10-24 & LAMOST LRS & 3700--9000 & 2.5 & 3$\times$1200 & 3.6 & 12 \\
2011-11-20 & LAMOST LRS & 3700--9000 & 2.5 & 2$\times$1200 & 3.0 & 8 \\
2024-09-15 & P200/DBSP-B1200 & 3850--5400 & 1.4 & 5$\times$1800 & 1.2 & 62 \\
           & P200/DBSP-R600  & 5750--9100 & 2.79 & 6$\times$1500 & 1.2 & 107 \\
\hline
\end{tabular}
\end{center}
\end{table*}

\section{Data}

The Large Sky Area Multi-Object Fiber Spectroscopic Telescope \citep[LAMOST;][]{Cui2012, Zhao2012}, equipped with 4000 robotic fibers and a 20\,deg$^2$ field of view, enables wide-field spectroscopic surveys across the entire extent of M31, spanning from its inner bulge to the outermost halo. Using LAMOST data, we recently published a catalog of supergiant candidates in M31, identified primarily based on their radial velocities \citep{Chen2025}. One of the notable objects from this catalog is J0048+4154. It was flagged due to prominent absorption features in its LAMOST spectrum. Moreover, its galactocentric distance (to the M31 center) of about 34\,kpc is well beyond the typical search radius ($<$25\,kpc), suggesting an unusual origin. These factors motivated deeper follow-up observations to confirm its nature. In this section, we describe the spectroscopic and photometric data used in our analysis, including both archival resources and new observations obtained for this study.

\subsection{Spectroscopic Data}

J0048+4154 was observed twice with LAMOST using its low-resolution mode ($R \sim 1800$), with nearly a one-month interval between the two epochs. To obtain higher-quality data, we conducted follow-up observations using the Double Spectrograph \citep[DBSP;][]{Oke1982} on the 200-inch Hale Telescope at Palomar Observatory (P200). Observations were carried out with the D55 dichroic, employing the 1200/5000 grating for the blue arm and the 600/10000 grating for the red arm. This setup delivered a spectral resolution of 1.4\,\AA\ (FWHM) over 3850--5400\,\AA\ in the blue and 2.79\,\AA\ (FWHM) over 5750--9100\,\AA\ in the red, excluding a gap between 6010--6325\,\AA. A 1\arcsec\ slit was used for all DBSP exposures. A summary of the spectroscopic observations is provided in Table~\ref{tab: observations}. The DBSP data were reduced using standard procedures with the \texttt{PypeIt} pipeline\footnote{\url{https://github.com/pypeit/PypeIt}} \citep{pypeit1, pypeit2}.

\subsection{Photometric Data}

To construct a reliable spectral energy distribution (SED) for J0048+4154, we compiled multi-wavelength photometric data spanning from optical to mid-infrared.  We adopted the optical $ugriz$-band photometry from the Sloan Digital Sky Survey \citep[SDSS;][]{sdss}, near-infrared $JHK_s$-band measurements from the Two Micron All-Sky Survey \citep[2MASS;][]{2mass}, and mid-infrared $W1W2$ data from the Wide-field Infrared Survey Explorer \citep[WISE;][]{allwise}. We exclude the $W3$- and $W4$-band magnitudes from WISE, as the source is undetected in these longer-wavelength images, rendering their photometry unreliable.

In addition, we retrieved high-cadence light curves from the Zwicky Transient Facility \citep[ZTF;][]{Bellm2019, Masci2019, Graham2019} via the NASA/IPAC Infrared Science Archive \citep[IRSA;][]{ipac}. The ZTF DR23 data set spans from March 2018 to October 2024, offering $\sim$7 years of monitoring in the $g$ and $r$ bands with a median cadence of 1--3 days. For our target with $r \sim 17$\,mag, the photometric precision is about 0.02\,mag.

\begin{figure*}
\centering 
    \includegraphics[width=2.12\columnwidth]{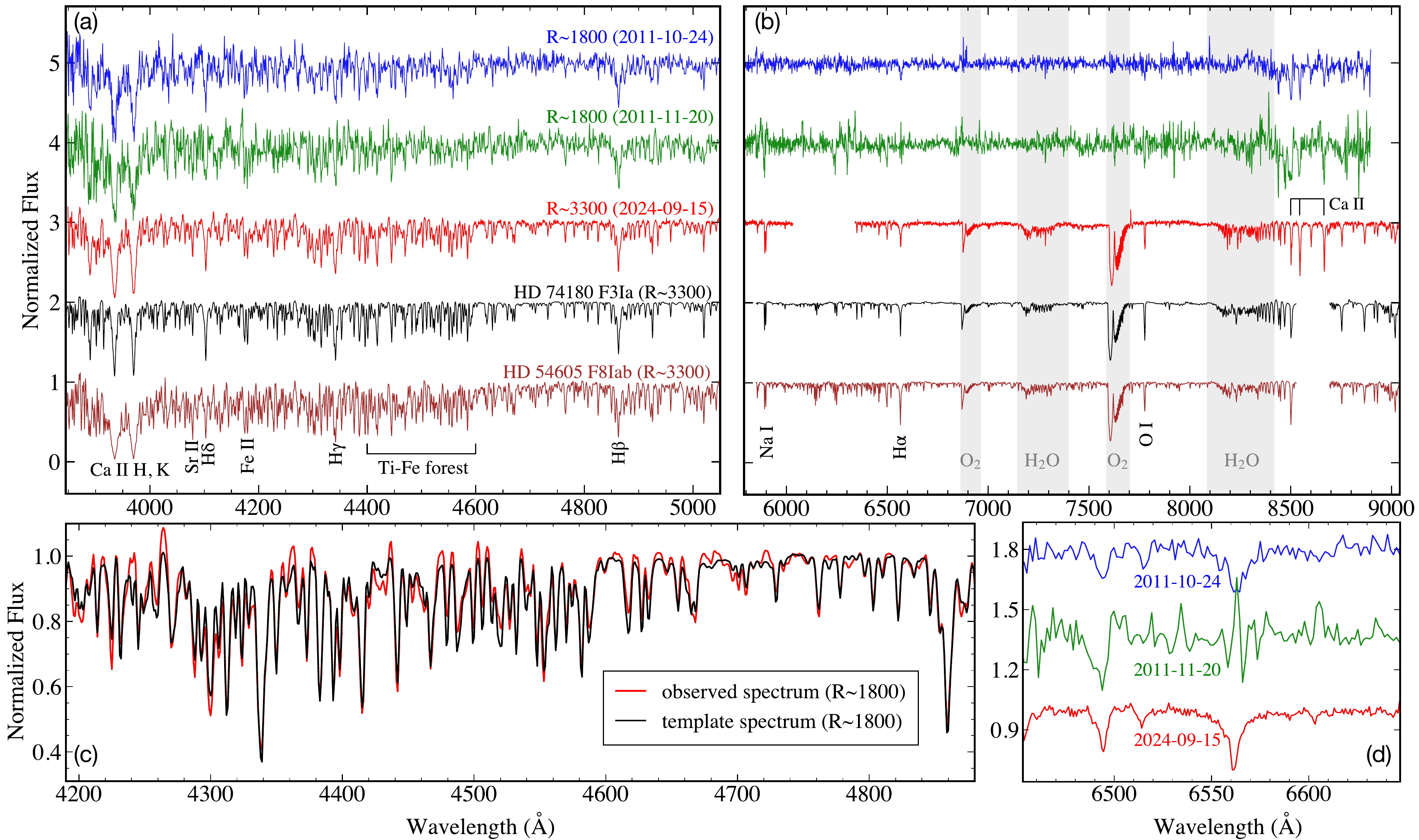}
    \caption{
    Normalized spectra of J0048+4154 compared to template supergiants. 
    (a) Blue spectra showing J0048+4154 (top three) alongside two UVES-POP spectra of Galactic F-type supergiants. Key absorption features are labeled, and radial velocity corrections have been applied.
    (b) Same as (a), but for the red spectra. Telluric features, which remain uncorrected in both the DBSP and UVES-POP spectra, are indicated by gray shading.
    (c) Comparison of the normalized DBSP spectrum of J0048+4154 (black) with a velocity-shifted template spectrum of HD\,74180 (red). 
    (d) Zoom-in on the H$\alpha$ region for J0048+4154. The color scheme matches panel (a), and no velocity corrections were applied.
    \label{fig: spectra}}
\end{figure*}

\section{Results}

\subsection{Spectral Characteristics and Radial Velocity} \label{subsec: spectral characteristics}

Fig.~\ref{fig: spectra} presents the normalized spectra of J0048+4154. For comparison, we included spectra of two Galactic F-type supergiants, HD\,74180 (F3\,Ia) and HD\,54605 (F8\,Iab), from the UVES-POP stellar spectral library \citep[Ultraviolet and Visual Echelle Spectrograph Paranal Observatory Project;][]{Bagnulo2003}. These templates were convolved to match the resolution of our DBSP spectrum using the {\tt laspec}\footnote{\url{https://github.com/hypergravity/laspec}} toolkit \citep{Zhang2020}. Similar to its Galactic counterparts, J0048+4154 shows strong, narrow absorption lines from ionized metal species. Notably, the \ion{Fe}{2} and \ion{Ti}{2} blends near $\lambda\lambda$4172--4178\,\AA\ and $\lambda\lambda$4395--4400\,\AA\ are prominent, indicating high luminosity. The \ion{Sr}{2} $\lambda$4077 line, when compared to nearby H$\delta$, provides a useful temperature diagnostic. The absence of the G band feature suggests an earlier spectral type than G. Based on these features, we classify J0048+4154 as an F5--F8 supergiant (F5--8\,I), with the relatively weak hydrogen lines favoring a later subtype near F8.

In the red spectral region, the \ion{O}{1} triplet at 7774\,\,\AA\ serves as a useful luminosity indicator for A- to F-type supergiants \citep{Osmer1972, Ferro2003}, and exhibits strong temperature dependence in G-type supergiants with $4500<T_{\text{eff}}<6000\,$K \citep{Kovtyukh2012}. We measured an equivalent width of $1.8 \pm 0.1$\,\AA\ for this feature in the DBSP spectrum using a Gaussian fit. This value significantly exceeds that expected for dwarfs or giants and is consistent with a supergiant classification. Although the resolution does not fully resolve the triplet, the measured strength corresponds to an absolute visual magnitude of $M_V \approx -7.8$\,mag based on the calibration of \citet{Ferro2003}, in agreement with results from broadband photometry (see Section~\ref{sec:photometric analysis and physical properties}).

The DBSP red spectrum shows some unusual features. The \ion{Na}{1} D lines are relatively strong, and H$\alpha$ appears weaker than expected. A comparison with LAMOST spectra suggests possible variability in the H$\alpha$ profile (Fig.~\ref{fig: spectra} (d)), including asymmetry and potential emission components. These signatures hint at weak H$\alpha$ emission. However, we note that the H$\alpha$ profile displayed in the LAMOST spectra could be caused by unphysical effects. In particular, the relatively large fiber diameter ($\sim$3\arcsec) and imperfect sky subtraction may have introduced contamination from M31's background starlight.

To measure radial velocities, we used the cross-correlation function (CCF) method following the approach of \citet{Zhang2021}. We used the UVES-POP spectrum of HD\,74180 as the primary template and matched all spectra to a common resolution of $R \sim 1800$. While J0048+4154 and HD\,74180 differ in physical properties, their spectral similarity ensures reliable velocity determination (Fig.~\ref{fig: spectra} (c)). The CCF was computed over the 4000--5250\,\AA\ range in 1\,\AA\ steps. We used \texttt{scipy.optimize.minimize} with the Nelder-Mead algorithm \citep{Nelder1965} to locate the velocity corresponding to the CCF maximum. To estimate uncertainties, we performed 300 Monte Carlo simulations by injecting Gaussian noise into the observed spectra. For consistency, we repeated the CCF analysis using the DBSP spectrum of J0048+4154 as the template. The results agree well with those obtained using HD\,74180. We also tested HD\,54605 as a template, which resulted in slightly lower CCF peaks and small systematic shifts ($\sim2\,\kms$), but did not affect the overall conclusions. No significant radial velocity changes were detected across the three epochs, within the measurement uncertainties. The results are summarized in Table~\ref{tab: velocity}.

\begin{table}
\caption{Radial Velocity Measurements\label{tab: velocity}}
\setlength{\tabcolsep}{5.pt}
\begin{center}
\begin{tabular}{cccc}
\hline
\hline
 & CCF Max$^a$ & $V_r$ ($\kms$)$^a$ & $\delta V_r$ ($\kms$)$^b$\\
\hline
LAMOST1 & 0.77 & $-203.7 \pm 3.8$ & $1.4 \pm 3.9$\\
LAMOST2 & 0.67 & $-200.7 \pm 5.4$ & $5.6 \pm 5.5$\\
DBSP    & 0.95 & $-206.9 \pm 0.3$ & $0.0$\\
\hline
\end{tabular}
\end{center}
\tablenotetext{a}{Maximum CCF values and Radial velocities derived using the UVES-POP spectrum of HD\,74180 as the template.}
\tablenotetext{b}{Radial velocity differences with the DBSP spectrum of J0048+4154 as the template.}
\end{table}

To assess whether J0048+4154 follows the rotation of M31's disk, we compared its velocity with the empirical kinematic model from \citet{Massey2016a}, which describes the expected radial velocity ($V_{\rm exp}$) for Red Supergiants (RSGs) as a function of position. At the location of J0048+4154, the extrapolated model under the assumption of a flat rotation curve, a fixed inclination angle, and a fixed position angle predicts $V_{\rm exp} = -196.8\,\kms$, which closely matches our measured velocity of $\sim-204\,\kms$. This agreement supports the idea that J0048+4154 formed in situ, participating in the rotational motion of M31's young stellar disk. We return to this point in Section~\ref{sec: environments and possible origin}.

\subsection{Photometric Analysis and Physical Properties} \label{sec:photometric analysis and physical properties}

\subsubsection{Spectral Energy Distribution} 
\label{subsubsec: spectral energy distribution}

\begin{figure}
\centering 
\includegraphics[width=0.95\columnwidth]{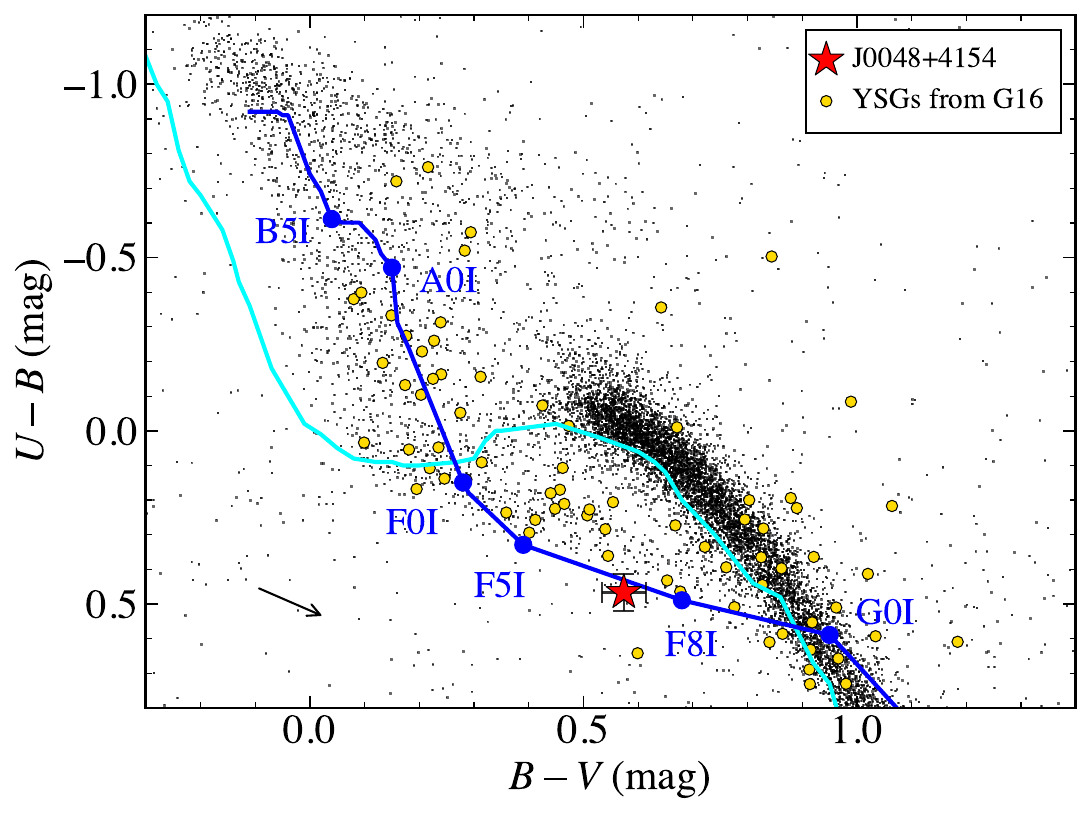}
\caption{
$U-B$ versus $B-V$ color-color diagram. Small black dots represent stars from the LGGS catalog \citep{Massey2006, Massey2016b} with $V \leq 19.5$\,mag, while yellow circles indicate spectroscopically confirmed YSGs from G16. The red asterisk marks the position of J0048+4154. The intrinsic color sequences of dwarf stars (cyan line) and Ia supergiants (dark blue line) from \citet{Fitzgerald1970} are shown, with the latter reddened by $E(B-V)=0.13$\,mag. The reddening vector, calculated by the extinction law from \citet{Cardelli1989} with $R_V=3.1$, is indicated by the black arrow in the lower left.
\label{fig: two_color}}
\end{figure}

We have converted the SDSS $ugriz$ magnitudes of J0048+4154 to the Johnson system using the transformation equations from \citet{Jester2005}, yielding $B-V = 0.57 \pm 0.04$\,mag and $U-B = 0.47 \pm 0.05$\,mag. Color-color diagrams provide useful diagnostics for identifying the intrinsic properties of supergiants. Fig.~\ref{fig: two_color} shows the position of J0048+4154 in the color-color diagram relative to the broader stellar population and known YSGs. The majority of sources from the Local Group Galaxies Survey \citep[LGGS;][]{Massey2006, Massey2016b} fall along the dwarf sequence, consistent with foreground stars from the Milky Way. We applied a reddening correction of $E(B-V) = 0.13$\,mag to the supergiant sequence, in line with the typical extinction toward OB stars in M31 as reported by \citet{Massey2007}.

J0048+4154 lies close to the reddened supergiant sequence, consistent with a spectral type between F5 and F8, and likely closer to F8. This result aligns well with our spectroscopic classification (Section~\ref{subsec: spectral characteristics}). Notably, its position also suggests lower extinction compared to other YSGs in \citet[][hereafter G16]{Gordon2016}, which is expected given its location in the outer regions of M31, where dust content is lower. 

To further constrain the physical properties of J0048+4154, we constructed its SED using archival photometry from SDSS, 2MASS, and WISE. The SED fitting was performed using the Python package {\sc speedyfit}\footnote{\url{https://github.com/vosjo/speedyfit}} \citep{Vos2017, Vos2018}, which applies a Markov chain Monte Carlo (MCMC) method to model single or binary stellar SEDs using atmosphere templates.

In this analysis, we adopted solar metallicity atmosphere models from \citet{Kurucz1993}. The resulting parameters are summarized in Table~\ref{tab: sed}. We note that surface gravity is poorly constrained from SED fitting alone, and the derived $\log g$ value should be treated with caution. In addition, two main degeneracies affect the fit: one between effective temperature and extinction, and another between stellar radius and distance. These degeneracies cannot be broken with the current dataset. Nevertheless, the derived parameters are broadly consistent with our spectroscopic results and are sufficiently robust for comparative purposes.

\begin{table}
\caption{Physical Parameters of J0048+4154 from SED Fitting.\label{tab: sed}}
\setlength{\tabcolsep}{20.pt}
\begin{center}
\begin{tabular}{lr}
\hline
\hline
Parameter & Value \\
\hline
$T_{\text{eff}}$ (K) & $6357_{-118}^{+121}$ \\
$\log g$ (dex) & $1.0_{-0.2}^{+0.2}$ \\
$R/R_{\odot}$ & $262_{-16}^{+17}$ \\
$E(B-V)$ (mag) & $0.12_{-0.03}^{+0.03}$ \\
Distance Modulus (mag) & $24.47_{-0.13}^{+0.14}$ \\
$\log L/L_{\odot}$ & $5.00_{-0.06}^{+0.06}$ \\
\hline
\end{tabular}
\end{center}
\tablecomments{Uncertainties represent 2$\sigma$ confidence intervals.}
\end{table}

\begin{figure}
\centering 
\includegraphics[width=1.0\columnwidth]{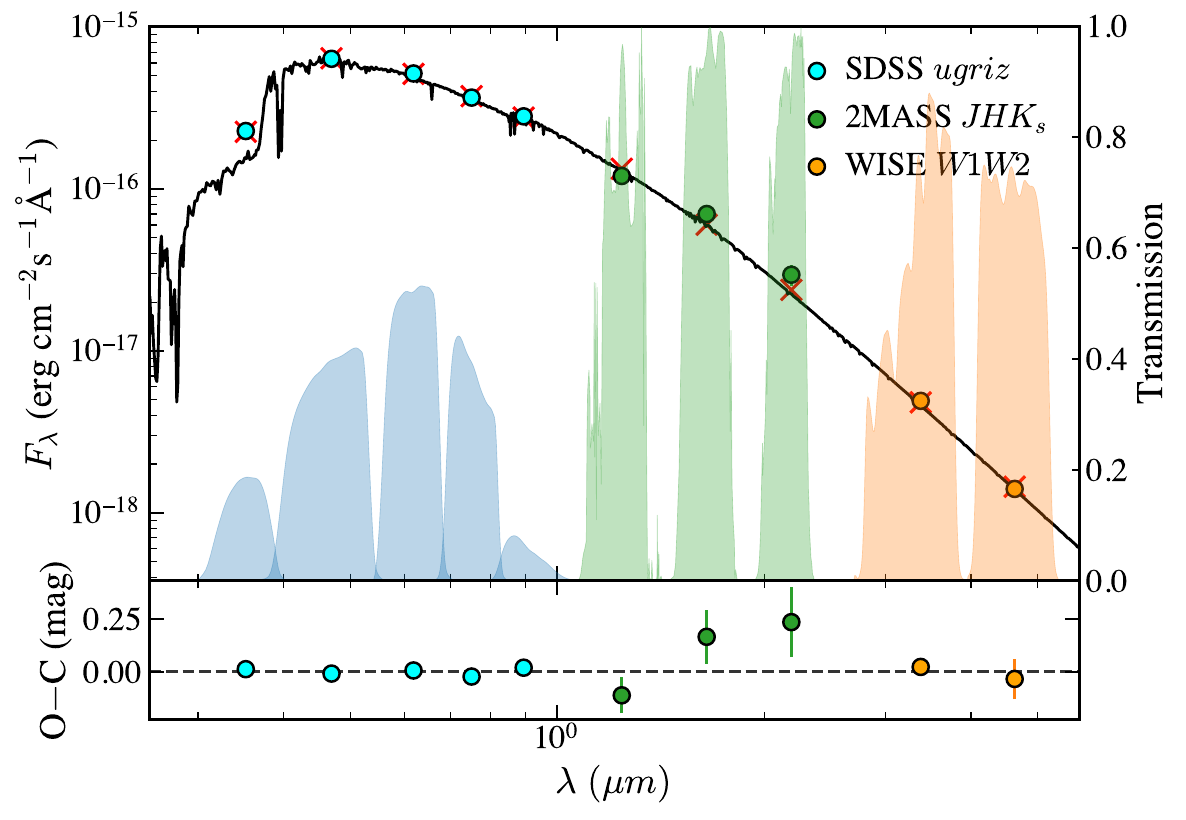}
\caption{
Spectral energy distribution (SED) fit for J0048+4154. Circles with error bars indicate observed fluxes from multiple surveys, color-coded by different filter set. Red crosses represent model-predicted fluxes integrated over the corresponding filter transmission curves (shaded regions). The black line shows the best-fit model.
\label{fig: sed}}
\end{figure}

Fig.~\ref{fig: sed} shows the best-fit SED model. No significant infrared excess is observed, indicating the likely absence of circumstellar dust produced by recent mass-loss activity. Visual inspection of WISE imaging confirms that the source is undetected in the $W3$ and $W4$ bands. Additionally, no excess is seen in the SDSS $u$-band, although the lack of ultraviolet data limits our ability to assess color excess at shorter wavelengths.

\subsubsection{Light Curves} \label{subsubsec: light curves}

\begin{figure}
\centering 
\includegraphics[width=1.0\columnwidth]{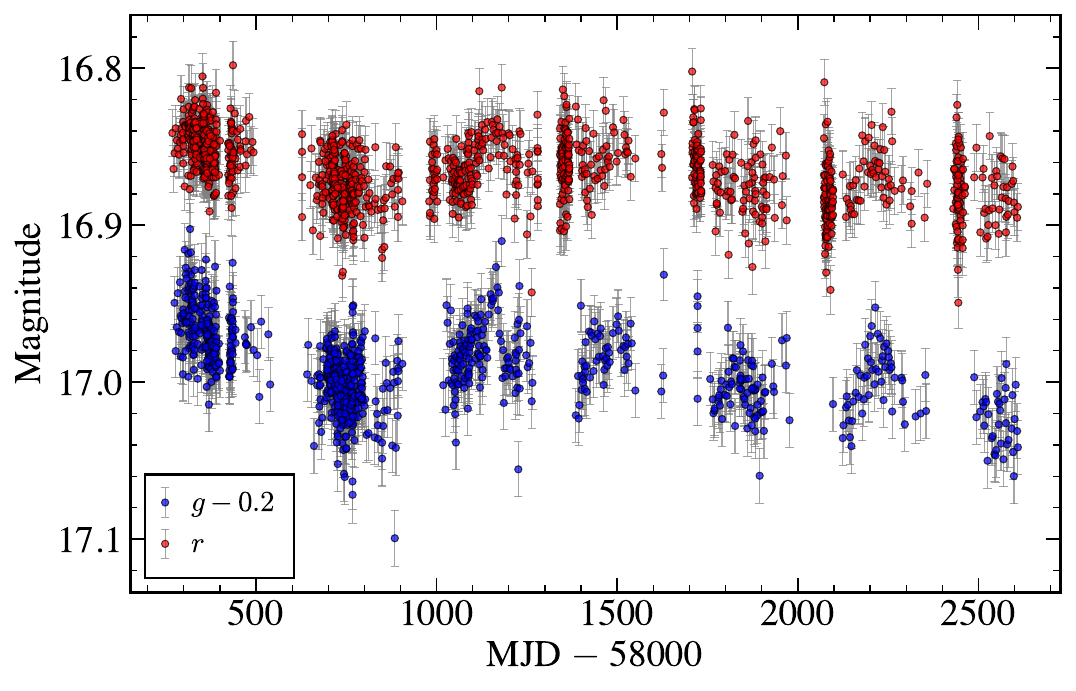}
\caption{
ZTF optical light curves of J0048+4154 in the $g$ and $r$ bands. Magnitudes are offset for clarity.
\label{fig: lc}}
\end{figure}

J0048+4154 lies near the upper boundary of the classical instability strip in the HR diagram \citep[e.g.,][]{Ekstrom2012, Yusof2022}, but it is hotter than typical high-luminosity Cepheids (see Fig.~\ref{fig: HR}). Fig.~\ref{fig: lc} shows its light curves in the $g$ and $r$ bands from ZTF. Over a seven-year baseline, the source exhibits relatively stable brightness, with low-amplitude variations on multiple timescales.

The mean magnitude in the ZTF $r$ band is 16.87\,mag, with a root-mean-square (rms) scatter of 0.02\,mag, which is comparable to the photometric precision at this brightness level. Using the Lomb-Scargle periodogram \citep{Lomb1976, Scargle1982, VanderPlas2018}, we identified weak peaks corresponding to possible periodicities between 200 and 1300 days. While this may suggest the presence of multiple pulsation modes, the current dataset does not allow for robust confirmation. Further high-precision photometric monitoring will be required to investigate potential variability in greater detail.

\subsubsection{Initial Mass and Evolutionary Status} \label{subsubsec: comparison with evolutionary tracks}

\begin{figure}
\centering 
\includegraphics[width=1.0\columnwidth]{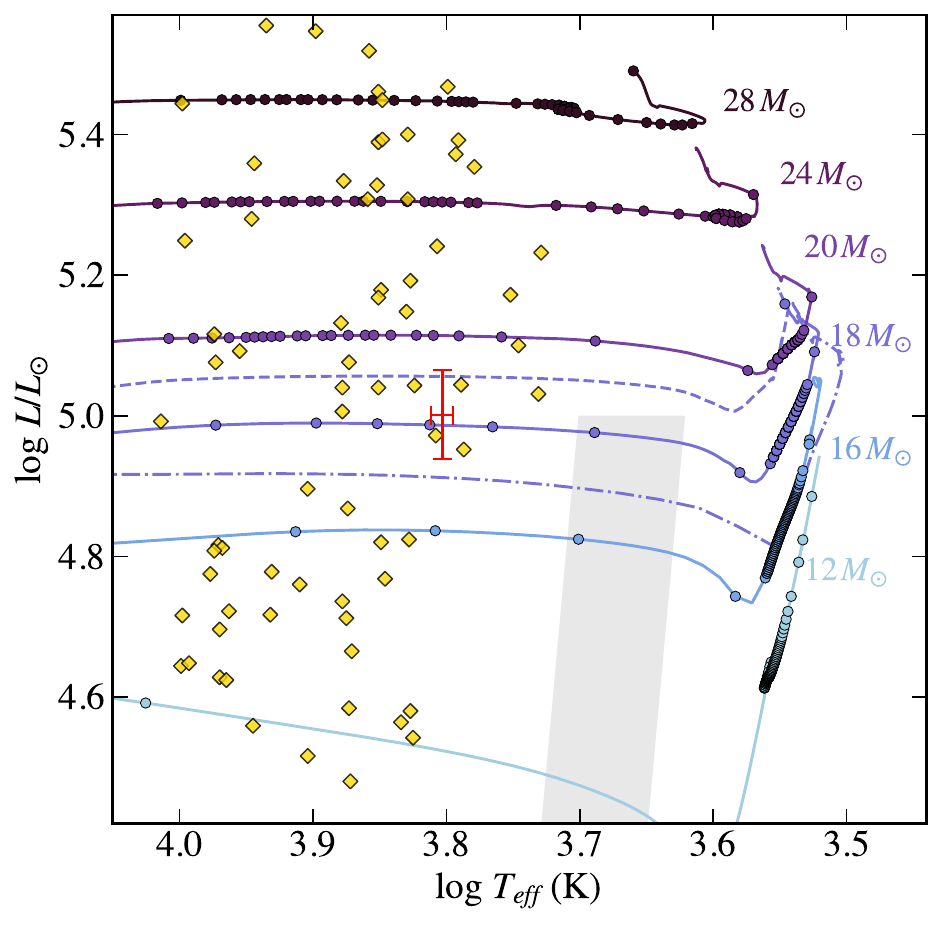}
\caption{
Hertzsprung-Russell diagram showing the position of J0048+4154 (red symbol with 2$\sigma$ error bars). Solid lines represent MIST evolutionary tracks \citep{Dotter2016, Choi2016} for stars with initial masses between 12 and $28\,M_{\odot}$ at solar metallicity ($Z=0.0142$), assuming an initial rotation of 40\% of the critical velocity. Colored dots mark evolutionary steps at $10^4$\,yr intervals. Dashed and dash-dotted lines show $18\,M_{\odot}$ tracks at sub-solar ($Z=0.008$) and super-solar ($Z=0.025$) metallicities, respectively. Spectroscopically confirmed YSGs in M31 from G16 are shown as yellow diamonds. The grey shaded region indicates the instability strip from \citet{Tammann2003}.
\label{fig: HR}}
\end{figure}

To estimate the initial mass and evolutionary state of J0048+4154, we compared its luminosity and effective temperature (derived from SED fitting) with MIST stellar evolutionary models \citep{Dotter2016, Choi2016}. Fig.~\ref{fig: HR} shows its location in the Hertzsprung-Russell (HR) diagram, overlaid with evolutionary tracks for stars with initial masses from 12 to $28\,M_{\odot}$ at solar metallicity ($Z = 0.0142$), assuming an initial rotation of $v/v_{\text{crit}} = 0.4$. The tracks are sampled at $10^4$\,yr intervals to indicate evolutionary timescales. For reference, we also include YSGs from G16, excluding three hypergiants with luminosities beyond the plotted range.

Assuming solar metallicity, J0048+4154 is most consistent with an initial mass of approximately $18\,M_{\odot}$ and an age of roughly 10\,Myr. Metallicity gradients in M31 suggest a possible decrease with increasing galactocentric radius. Extrapolating the metallicity profile of \citet{Liu2022}, who studied blue supergiants out to $R_{\text{GC}} \approx 23.3$\,kpc, implies that J0048+4154 (at $\sim$34\,kpc) may have a metallicity closer to that of the Large Magellanic Cloud ($Z \approx 0.008$), though this estimate carries significant uncertainty. To assess the impact of metallicity, we also show $18\,M_{\odot}$ tracks at sub-solar and super-solar metallicities in Fig.~\ref{fig: HR}. The inferred initial mass varies by about $\pm 1\,M_{\odot}$ across this range, comparable to the uncertainty arising from the SED-derived luminosity. An additional uncertainty of $\sim\pm 1\,M_{\odot}$ arises from differences between stellar evolution models, such as those from the Geneva group \citep{Ekstrom2012}.

The position of J0048+4154 in the HR diagram suggests it is in a rapid evolutionary phase. Under single-star evolution, such a star is expected to be crossing the Hertzsprung gap on its way from the blue to red supergiant phase, ultimately ending its life as a hydrogen-rich core-collapse supernova \citep[e.g.,][]{Ekstrom2012}. However, an alternative view posits that the star may be evolving blueward after a RSG phase. Models that include substantial mass loss during the RSG phase predict a return to hotter temperatures in the HR diagram before the terminal explosion \citep[e.g.,][]{Ekstrom2012, Georgy2012, Meynet2015}. 

G16 found that approximately 40\% of YSGs in M31 may be post-RSG objects, typically with luminosities above $10^5\,L_{\odot}$ and evidence for dusty mass loss. While J0048+4154 lies near that luminosity threshold, its SED does not show infrared excess (Section~\ref{subsubsec: spectral energy distribution}), suggesting little or no circumstellar dust. Recently, a new class of fast yellow pulsating supergiants has been proposed as tracers of the post-RSG phase \citep{Dorn-Wallenstein2022}. These stars also appear to lie near a luminosity boundary around $10^5\,L_{\odot}$. Future investigation of high-frequency variability in J0048+4154 could help clarify its nature, although such studies are observationally challenging due to the target's faintness and potential contamination from nearby stars \citep{Pedersen2023}.

While our analysis supports the classification of J0048+4154 as a massive YSG, we briefly consider whether it could be a lower-mass evolved star. Super-asymptotic giant branch (super-AGB) stars, with initial masses of 6--12\,$M_{\odot}$, are predicted to reach luminosities up to $\sim$$10^5\,L_{\odot}$ \citep{Doherty2017}. However, these stars typically have cooler effective temperatures ($T_{\text{eff}} < 4000$\,K) and are often enshrouded in dusty shells due to strong mass loss, producing noticeable infrared excess. Neither condition is observed in J0048+4154, effectively ruling out an AGB or post-AGB origin.

\begin{figure*}
\centering 
\includegraphics[width=2.1\columnwidth]{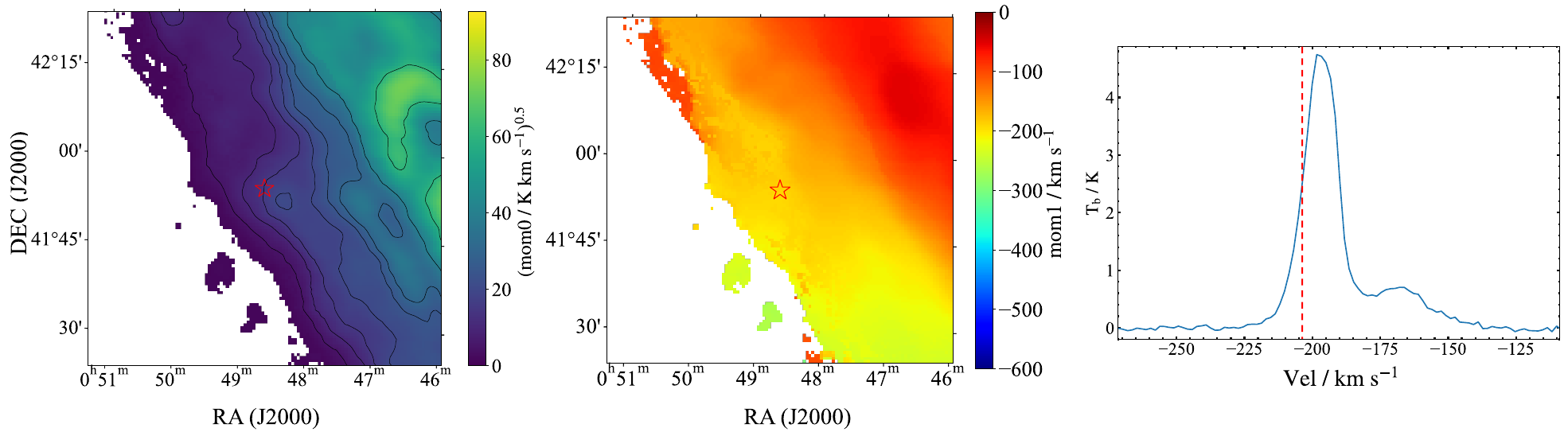}
\caption{
Left: Integrated H\,\textsc{i} brightness
temperature (moment 0) map of the region surrounding J0048+4154, shown on a square-root scale. Contours indicate levels of [5, 10, 15, 20, 25, 30, 40, and 60]$^2$\,K\,km\,s$^{-1}$. The spatial resolution is $\sim$3.4\arcmin, and the position of J0048+4154 is marked by a red asterisk. 
Middle: Same as left, but showing the flux-weighted velocity (moment 1) map. 
Right: H\,\textsc{i} spectrum at the location of J0048+4154. The average radial velocity of J0048+4154 ($\sim -204\, \kms$) is indicated by a red dashed line.
All data are from FAST observations (Zhang et al. 2025, in prep.).
\label{fig: HI}}
\end{figure*}

\begin{figure}
\centering 
\includegraphics[width=1.0\columnwidth]{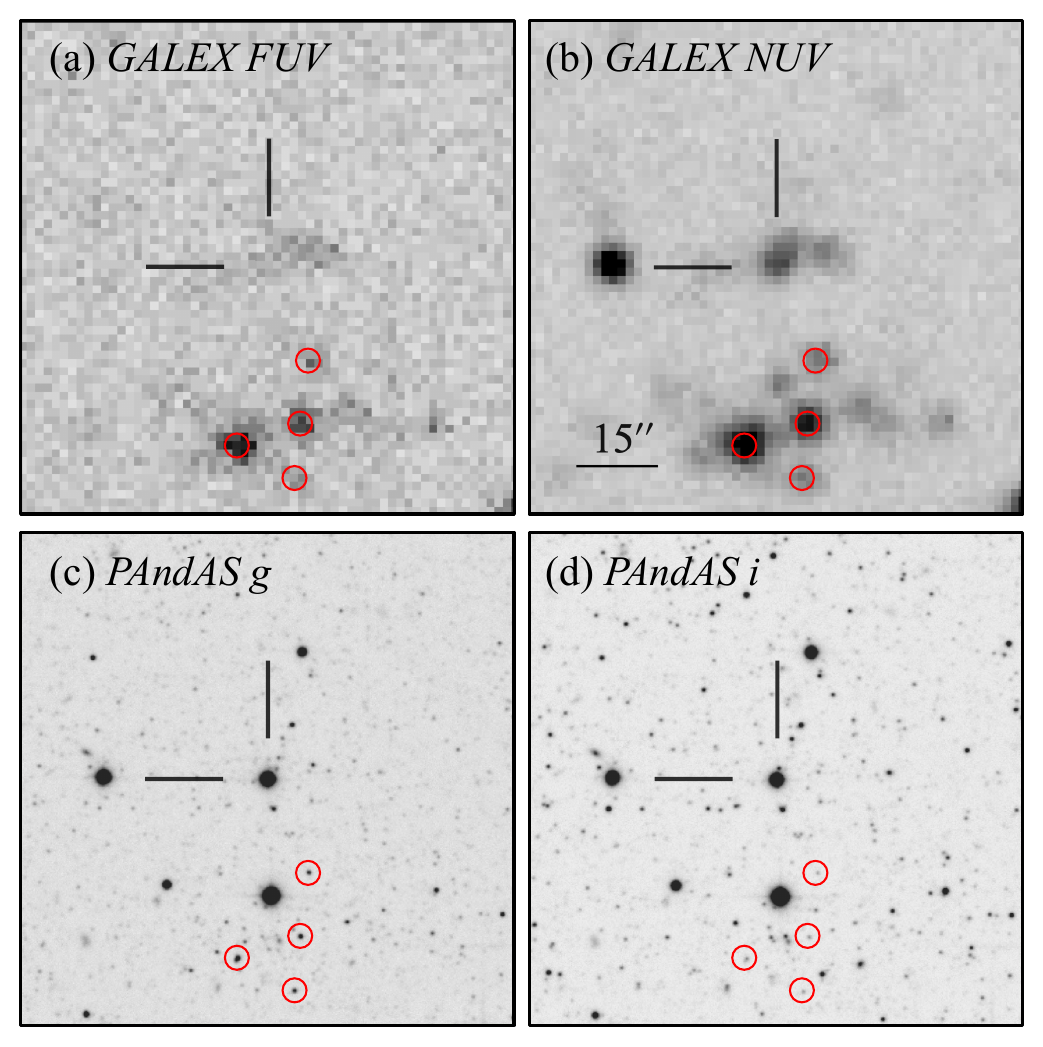}
\caption{
Multiwavelength images of J0048+4154. Four nearby UV-bright stars are marked with red circles. The scale bar of 15\arcsec\ corresponds a projected distance of 57\,pc. Each panel is 1.5\arcmin\ across, with north up and east to the left.
\label{fig: multiband}}
\end{figure}

\begin{table*}
\caption{Photometry of UV-bright Stars Near J0048+4154\label{tab: photometry}}
\setlength{\tabcolsep}{10.pt}
\begin{center}
\begin{tabular}{lcccccc}
\hline
\hline
SDSS ID & Distance$^a$ & F148W$^b$ & $g$ (SDSS) & $U-B^c$ & $B-V^c$ & Q-index \\
\hline
J004838.15+415341.9 & 380\,pc& $22.38 \pm 0.16$ & $22.62 \pm 0.13$ & \dots & \dots & \dots \\
J004837.92+415403.7 & 109\,pc & $22.73 \pm 0.20$ & $22.58 \pm 0.12$ & \dots & \dots & \dots \\
J004838.07+415352.1 & 255\,pc & $22.36 \pm 0.18$ & $21.92 \pm 0.07$ & $-0.82 \pm 0.15$ & $-0.04 \pm 0.14$ & $-0.79 \pm 0.18$ \\
J004839.10+415348.0 & 446\,pc & $21.89 \pm 0.14$ & $21.56 \pm 0.05$ & $-0.73 \pm 0.13$ & $-0.23 \pm 0.12$ & $-0.56 \pm 0.16$ \\
\hline
\end{tabular}
\end{center}
\tablenotetext{a}{Deprojected distances to J0048+4154.}
\tablenotetext{b}{F148W magnitudes are taken from the AstroSat/UVIT compact source catalog \citep{Leahy2025}, except for J004838.07+415352.1, which uses the earlier release \citep{Leahy2020}.}
\tablenotetext{c}{Colors converted from SDSS photometry using relations from \citet{Jester2005}.}
\end{table*}

\section{Possible Origin of J0048+4154} \label{sec: environments and possible origin}
\subsection{The H\,\textsc{i} External Arm} \label{subsec: the external arm in HI}

M31 hosts a large, warped H\,\textsc{i} disk, with several extended and asymmetric structures identified in its outskirts \citep[e.g.,][]{Carignan2006, Chemin2009, Braun2009, Corbelli2010}. Among these is the so-called external arm located in the southeastern edge of the galaxy \citep{Chemin2009}. This structure is particularly faint and lacks detectable emission from the ultraviolet to the far-infrared, making it one of the most elusive components of M31's outer disk. Notably, the position of J0048+4154 aligns closely with this external arm.

To investigate this association, we used deep H\,\textsc{i} observations from the Five-hundred-meter Aperture Spherical Radio Telescope \citep[FAST;][]{Nan2011}, which offers high sensitivity ideal for probing the diffuse gas in M31's outskirts. A more detailed analysis of these observations will be presented in Zhang et al. (in prep.).

The left panel of Fig.~\ref{fig: HI} shows the integrated H\,\textsc{i} brightness temperature map around J0048+4154. The star lies near the edge of the H\,\textsc{i} disk in a region of relatively low gas density. The mapped region corresponds to a zoomed-in view of the external arm; for a broader context, see Fig.~5 in \citet{Chemin2009}. FAST data confirm that this faint, arm-like feature extends beyond a galactocentric radius of 30\,kpc. It appears to connect with M31's southwestern spiral arm at one end, while gradually fading in the northeast without a clear boundary.

The morphology and kinematics of the external arm are unusual. It shows a higher inclination and distinct position angle compared to the inner disk, and occupies a unique region in position-velocity space. Moreover, the absence of a corresponding structure on the opposite side of the disk suggests that it may have originated from an asymmetric external perturbation. \citet{Chemin2009} proposed that this external arm may have been shaped by tidal interactions with low-mass satellites such as NGC\,205. If this is the case, J0048+4154, given its short lifetime, may represent a rare example of massive star formation triggered by such external interactions in the outer disk.

The kinematic evidence also supports this association. As shown in the middle and right panels of Fig.~\ref{fig: HI}, the H\,\textsc{i} velocity field is warped in this region. The spectrum at the location of J0048+4154 reveals two components: a dominant peak at $\sim$$-200$\,km\,s$^{-1}$ and a secondary component at $\sim$$-165$\,km\,s$^{-1}$. The primary component closely matches the stellar radial velocity of J0048+4154 ($\sim -204$\,km\,s$^{-1}$), strongly suggesting a physical connection to the external arm, both spatially and kinematically.

\subsection{Local Environment} \label{subsec: local environments}
The V-band surface brightness at the location of J0048+4154 is estimated to be approximately $27-28$ mag\,arcsec$^{-2}$ \citep{Tempel2011, Courteau2011, Ibata2014}, albeit with considerable uncertainty. This exceptionally faint value indicates that the star resides in a region of remarkably sparse stellar density. To better understand the immediate environment of J0048+4154, we examined ultraviolet and optical images of the region (Fig.~\ref{fig: multiband}). These include $FUV$ and $NUV$ images from GALEX \citep{Morrissey2007} and $g$- and $i$-band images from the Pan-Andromeda Archaeological Survey \citep[PAndAS;][]{McConnachie2009, McConnachie2018}. In the deep optical images, J0048+4154 appears as a point source. Infrared imaging from 2MASS and WISE (not shown) also reveals no extended or composite structure. These observations suggest that J0048+4154 is isolated, and not embedded in an H\,\textsc{ii} region or part of a young stellar cluster, unlike many massive stars in M31's inner disk. 

Although J0048+4154 is faint in the ultraviolet and lacks precise GALEX photometry, the surrounding field shows several nearby blue sources. We identified a group of UV-bright stars located just south of the target. Four of these stars are cataloged in the recent AstroSat/UVIT survey of M31 \citep{Leahy2020, Leahy2025}. Their positions are marked in Fig.~\ref{fig: multiband}, and their photometric properties are listed in Table~\ref{tab: photometry}. For the two brighter stars with reliable photometries, we converted their SDSS colors to Johnson $U\!-\!B$ and $B\!-\!V$ using the transformations from \citet{Jester2005}, valid for sources with $U\!-\!B < 0$. We then calculated the reddening-free Q-index as $Q = (U\!-\!B) - 0.72\,(B\!-\!V)$ \citep{Johnson1953, Massey1998}. A Q-index of $Q < -0.6$ typically corresponds to early B-type main-sequence stars or B-type supergiants \citep{Massey1998, Massey2006}. Based on their photometric properties at the distance of M31, these nearby UV-bright stars are likely early B-type main-sequence stars.

The extent of this four star group is about 350\,pc, assuming they lie within the disk plane. The presence of additional faint, undetected sources in the field suggests a potentially larger structure. While this size exceeds that of typical open clusters, it remains unclear whether these stars form a gravitationally bound association or a more loosely connected stellar group. Notably, we did not identify any other UV-bright stellar concentrations, star-forming regions, or known H\,\textsc{ii} regions in the surrounding area at comparable galactocentric distances. This reinforces the idea that J0048+4154 may have formed in a sparsely populated and dynamically perturbed region of M31's outer disk.

\subsection{Formed in Situ or Migrated?}

J0048+4154 currently represents the most distant known supergiant in M31, raising the question of whether J0048+4154 truly formed locally or was instead displaced from a denser region within the galaxy. In recent years, several evolved runaway supergiants have been identified based on their unusual radial velocities \citep{Evans2015, Neugent2018, Comeron2020}. This prompts consideration of whether J0048+4154 might also be a runaway star, ejected from the outer edge of M31's inner star-forming regions.

However, the evidence argues against this scenario. J0048+4154's radial velocity is fully consistent with the expected local value from M31's rotation curve (see Section~\ref{subsec: spectral characteristics}) and closely matches the velocity of surrounding H\,\textsc{i} gas. This alignment indicates minimal motion relative to its local environment along the line of sight. Furthermore, even if a significant tangential velocity were present, its  trajectory would not intersect with the inner disk where active star formation is concentrated. These factors strongly favor an in situ origin or at least formation in close proximity to the star's current location.

One possibility is that J0048+4154 originated from the nearby group of young stars to its south and was displaced by a modest velocity kick, possibly from binary interactions. Assuming a relative velocity of $\sim$30\,km\,s$^{-1}$ and a travel time of 5\,Myr, the star could have moved about 150\,pc, which is greater than its current deprojected separation of $\sim$109\,pc from SDSS J004837.92+415403.7 (see Table~\ref{tab: photometry}). Spectroscopic measurements of radial velocities for these neighboring young stars would help test this hypothesis.

Massive stars such as supergiants are generally thought to form in giant molecular clouds, where the stellar initial mass function is sampled \citep{Lada2003}. If J0048+4154 was formed locally, this implies that dense molecular clouds, perhaps smaller or transient, once existed in this part of M31's disk. Currently, most known molecular clouds in M31 are concentrated in the inner disk, particularly along the prominent ``Ring of Fire''. The detection of J0048+4154 in the outer disk highlights the importance of extending molecular gas searches to larger radii.

The discovery also suggests that M31 may share properties with extended ultraviolet (XUV) disk galaxies \citep{Thilker2007}, which are characterized by active star formation in low-density H\,\textsc{i} regions well beyond their optical radius. J0048+4154 may provide a rare example of such activity in M31's outskirts.

\section{Summary} \label{sec: summary and future perspective}

We have reported the discovery of J0048+4154, a massive YSG identified in the LAMOST spectroscopic survey. Located at a galactocentric distance of $\sim$34\,kpc, it is the most remote supergiant known in M31. Based on spectroscopic and photometric analyses, we classify this object as a late-F supergiant (F5--F8\,I). SED fitting yields an effective temperature of $6357^{+121}_{-118}$\,K and a luminosity of $\log L/L_{\odot} = 5.00^{+0.06}_{-0.06}$. Comparison with MIST evolutionary tracks suggests an initial mass of $\sim$18\,$M_{\odot}$ and an age of $\sim$10\,Myr. 

Crucially, J0048+4154 is spatially and kinematically associated with a faint H\,\textsc{i} arm in the warped extremities of M31's outer disk. This alignment strongly supports an in situ formation scenario, reinforcing the emerging picture that massive star formation can occur, albeit sporadically, well beyond traditionally recognized star-forming regions. The co-existence of UV-bright stars in the same vicinity further corroborates the presence of recent star formation in this sparsely populated environment.

The discovery challenges long-held perceptions that only high-density inner disks can generate massive stars, showcasing that outer disk environments, though diffuse, can occasionally meet the conditions necessary for massive star formation. These results offer an observational benchmark for testing models of gas accretion and warp-induced compression in galaxy outskirts. They also provide a real-life counterexample to simulations that struggle to reproduce extended star-forming disks. High-resolution spectroscopy of J0048+4154 will enable us to probe heavy elemental abundances in a region of M31 largely unexplored in chemical terms, shedding light on how outer disk stars may contribute to global enrichment processes. In summary, this work opens avenues for broader investigations into the high-mass star formation process in galaxy outskirts, offering a new lens through which to interpret the life cycle of spirals like M31 and beyond.

\section*{Acknowledgments}
We are grateful to the anonymous referee, whose excellent comments and suggestions significantly improved this article. This work is partially supported by the National Natural Science Foundation of China 12173034 and 12322304, the National Natural Science Foundation of Yunnan Province 202301AV070002 and the Xingdian talent support program of Yunnan Province. We acknowledge the science research grants from the China Manned Space Project with NO.\,CMS-CSST-2021-A09, CMS-CSST-2021-A08 and CMS-CSST-2021-B03. 

Guoshoujing Telescope (the Large Sky Area Multi-Object Fiber Spectroscopic Telescope LAMOST) is a National Major Scientific Project built by the Chinese Academy of Sciences. Funding for the project has been provided by the National Development and Reform Commission. LAMOST is operated and managed by the National Astronomical Observatories, Chinese Academy of Sciences.

This research uses data obtained through the Telescope Access Program (TAP), which is funded by the National Astronomical Observatories, Chinese Academy of Sciences, and the Special Fund for Astronomy from the Ministry of Finance. Observations obtained with the Hale Telescope at Palomar Observatory were obtained as part of an agreement between the National Astronomical Observatories, Chinese Academy of Sciences, and the California Institute of Technology. We are grateful to the staffs of Palomar Observatory for assistance with the observations and data management.

This work has used the data from the Five-hundred-meter Aperture Spherical radio Telescope (FAST).  FAST is a Chinese national mega-science facility, operated by the National Astronomical Observatories of Chinese Academy of Sciences (NAOC).

\bibliography{sample631}{}
\bibliographystyle{aasjournal}



\end{document}